\documentclass[11pt]{article}

\usepackage{amsmath}
\usepackage{amsfonts}
\usepackage{amssymb}
\usepackage{lscape}
\usepackage[T1]{fontenc}
\usepackage[latin1]{inputenc}
\usepackage{geometry}
\usepackage{comment} 
\begin{document}

\title{A Survey on Temporal Logics}

\author{Savas Konur\\Department of Computer Science, University of Liverpool}

\date{}

\maketitle

\begin{abstract}
This paper surveys main and recent studies on temporal logics in a broad sense by presenting various logic systems, 
dealing with various time structures, and discussing important features, such as decidability (or undecidability) results, 
expressiveness and proof systems.
\end{abstract}

\section{Introduction}

Temporal logics are formal frameworks which describe statements whose truth values change over time. 
Despite the fact that classical logics do not include time element, temporal logics
characterize state changes which depend on time. This makes temporal logics a 
richer notation than classical logics. 

Temporal logics can be considered as extensions of classical propositional and first-order logic. In fact, propositional temporal logics are an extension of propositional logic with temporal operators. Similarly, first-order temporal logics are extension of first-order logic with temporal modalities. Temporal logics are also special type of modal logics, where statements are evaluated on `worlds' which represent time instants. 

Although various aspects of time and logic have been studied, an up-to date comprehensive analysis of logic of time
does not exist in the literature. Some surveys (such as \cite{Ost92,CH04,OH95,GMS04}) can be found
in the literature  but these mainly concentrate on specific formal systems over specific structures of time; therefore,
they do not contain a broad analysis. The aim of this paper is to
outline main and recent developments in the field in a broad sense
by presenting various formal systems dealing with various time structures,
and discussing important features, such as (un)decidability results,
expressiveness and axiomatization systems. 

Temporal formalisms we will analyse include propositional/first-order linear temporal logics, branching temporal logics, partial-order temporal logics and interval temporal logics. We will summarize important results on decidability, axiomatizability, expressiveness, model checking, etc. for each logic analysed. We also provide a comparison of features of the temporal logics discussed.  

Note that in some instances we think it is more convenient to refer to the original text for clarification purposes. In the following,  we will use quotation marks to use the text from the original resources.

\section{Preliminaries} \label{sec:Classification-of-Temporal}

We can classify temporal logics based on several criteria. The common dimensions are `propositional versus first-order', `point-based versus interval-based', `linear versus branching', `discrete versus continuous', etc \cite{Eme95,Ven98,BMN00}. Below we discuss the most important  criteria to classify temporal logics.

\paragraph{Point versus interval structures:} There are two structure types to model time in a  temporal logic: \emph{points (instants)} and \emph{intervals}. A point structure $\mathcal{T}$  can be represented as $\left\langle T,<\right\rangle$, where $T$ is a nonempty time points, and $<$ is a `precedence' relation on $T$. Different temporal relationships can be described using different modal operators. Some logics include modal operators  which can express quantification over time. However, a relationship between intervals is difficult to express using a point-based temporal logic \cite{FM94}. 

Interval temporal logics are expressive, since these logics can express a relationship between two events, which are represented by intervals. Also, interval logics \cite{SMS82,SMV83,Mos83,Lad87,MS87,RG89,HS91} have a simpler and neater syntax to define a relationship between intervals, which provides a higher level abstraction than a point-based logic when modeling a system. This makes interval logic formulas much simpler and more comprehensive than point-based logic formulas. 

Some of the known interval operators are \emph{meets}, \emph{before}, \emph{during} \cite{All83}, which denote the ordering of intervals; \emph{chop} modality \cite{Ven91}, which denotes combining two intervals; and \emph{duration}, which denotes a length of an interval \cite{CH04}.

Interval structures can be considered in two ways: \emph{(i)} intervals are `primitive' objects \emph{(ii)} intervals are composed from points. \cite{vB91,MSV02,Vit05} consider intervals as primitive objects of time. \cite{vB91} defines  a `period structure' as the tuple $\left\langle \mathcal{I},\subseteq,\prec\right\rangle $, where $\mathcal{I}$ is a non-empty set of intervals, $\subseteq$ is a sub-interval relation, and $\prec$ is a precedence relation.
One particular problem of this approach is that theoretical analyses are usually very difficult. Also, although it is very easy to define properties  \emph{linearity}, \emph{density}, \emph{discreteness}, \emph{unboundedness} in a point-based logic, it is very difficult to define these properties in an interval logic where intervals are primitive objects. 

\cite{GMS04,HS91,Ven91} consider intervals as set of points, where the time flow is assumed as ``a strict partial ordering of time points''. Namely, an interval structure is defined as $\left\langle \mathcal{T},\mathcal{I}(\mathcal{T})\right\rangle$, where $\mathcal{T}=\left\langle T,<\right\rangle $ is a strict partial ordering and $\mathcal{I}(\mathcal{T})$ is a set of intervals. The properties mentioned above can be defined in an interval logic where intervals are composed of time instants. 

We conclude this section with the historical development of interval-based temporal logics. The concept of time intervals was first studied by Walker \cite{Wal47}. Walker considered a non-empty set of intervals, which is partially orderd. However, his work does not cover aspects of temporal logic in a general sense. In \cite{Ham71}  philosophical aspects of an interval ontology was analysed. In \cite{Hum79} an interval tense logic was introduced. 
 \cite{Dow79,Kam79,Rop80,Bur82,vB83,Gal84,Sim87} studied interval logics within the natural language domain. 
 It was argued that interval-based semantics are more convenient for human language and reasoning, and 
interval-based approach is more suitable than point-based approach for temporal constructions of natural language.
\cite{All83,All84,AH89,AF94} studied event relations and interval ordering. The authors introduced so-called Allen's thirteen interval relations and worked on axiomatisation and representation of interval structures.  Some further works on Allen's algebra were carried out by \cite{Lad87,Gal90}. Recently, \cite{RB06} investigated the relation between Allen's logic and LTL. Interval based-logics have been also applied to other fields in computer science. \cite{Par78,Pra79,HPS83} worked on process logic, where intervals are used as representation of information. Another important work was the development of \emph{interval temporal logic} \emph{(ITL)}, and its application to design of hardware components \cite{Mos83,HMM83}. Since the development of  ITL, various variations have been proposed so far. In particular, Duration Calculus \cite{CH04} is an extension of interval temporal logic with ``a calculus to specify and reason about properties of state durations''.

\paragraph{Temporal Structure:} There are important properties regarding the time flow and temporal domain structure. Some properties are summarized below:

Assume  $\left\langle T,<\right\rangle$ represents a temporal structure, where $T$ is a nonempty time points, and $<$ is a `precedence' relation on $T$. In a temporal logic the structure of time is \emph{linear} if any two points can be compared. Mathematically, a strict partial ordering is called linear if any two distinct points satisfy the condition: $\forall x,y:x<y\vee x=y\vee x>y$. This definition suggests that in linear temporal logics each time point is followed by only one successor point. 


Another class is the \emph{branching-time structures}, where the underlying temporal structure is  branching-like, and each point may have more than one successor points. The structure of time can be considered as a tree. A \emph{tree} is a set of time points $T$ ordered by a binary relation $<$ which satisfies the following requirements \cite{GHR94}:

\begin{itemize}
\item \noindent $\left\langle T,<\right\rangle $ is irreflexive;
\item \noindent $\left\langle T,<\right\rangle $ is transitive;
\item \noindent $\forall t,u,v\in T$ $u<t$ and $v<t \rightarrow$
$u<v, u=v$ or $u>v$ (i.e. the past of any point is linear);
\item \noindent $\forall x,y\in T, \exists z\in T$ such that $z<x$
and $z<y$ (i.e. $\left\langle T,<\right\rangle $ is connected).
\end{itemize}

One important characteristics of branching logics is that the syntax of these logics include path quantification which allows formulas to be evaluated over paths. However, linear temporal logics are restricted to only one path.

A temporal domain is \emph{discrete} with respect to the precedence relation $<$ if each non-final point is followed by a successor point. This can be formulated as follows: $\forall x,y$ ($x<y \rightarrow \exists z$ ($x<z \wedge \neg\exists w (x<w \wedge w<z$))) \cite{Spr02}. Majority of temporal logics used for system specification are defined on discrete time, where points represent system states. A state sequence, as a result of a program execution, can be considered as isomorphic to discrete series of positive integers. 

A temporal domain is \emph{dense} if, between any two distinct points, there is another point. This can be formally denoted  $\forall x,y (x<y \rightarrow \exists z (x<z<y$)) \cite{Spr02}. Above we mentioned that flow of discrete time can be represented as positive integers. Similarly, density can be represented as real numbers. It is noteworthy to mention that there is a distinction between \emph{density} and \emph{continuity}: ``A model of dense time is isomorphic to a dense series of rational numbers, meaning that there is always a rational number between any two rational numbers; whereas a model of \emph{continuous} time is isomorphic to a continuous series of real numbers'' \cite{Ven98}.

A temporal domain is \emph{bounded above} (\emph{bounded below}) if the temporal domain is bounded in the future (past) time. This can be formulated as follows: $\exists x \neg\exists y (x<y$ ($\exists x \neg\exists y (y<x$))) \cite{Spr02}. Similarly, a temporal domain is \emph{unbounded above} (\emph{unbounded below}) if each point has a successor (predecessor) point, which is formally denoted  $\forall x \exists y (x<y$  ($\forall x \exists y (y<x$))) \cite{Spr02}. 

A temporal domain \emph{is Dedekind complete} if all time point sets (non-empty) are bounded above, and they have a least upper bound.

Based on differences in temporal domain properties logics have different characteristics. For example, we can consider a temporal domain which is linear or branched; discrete or dense; finite/infinite in future and/or past, etc. All these choices result in different syntax, semantics, decidability and complexity. 

\section{Propositional Temporal Logics}

An important success in temporal logic study was the introduction of the temporal operators into the classical logic \cite{Kam68}. In \cite{Pnu77} Pnueli introduced a very influental \emph{Linear Temporal Logic (LTL)}. LTL can express properties of linear sequences of states. For example, properties such as `\emph{p} holds at some state in the sequence' or `\emph{p} holds at two consecutive states in the future' can be expressed in LTL. In \cite{SC85} Sistla and Clarke proved that the satisfiability and model checking problems of LTL are PSPACE-complete. \cite{SC85} shows that if the syntax is restricted only to $\diamondsuit$ (`sometime') operator, or  $X$ (`next') operator, then the complexity of the satisfiability problem reduces to NP-complete. However, when both operators are included in the syntax PSPACE-completeness is preserved. 

Recently, a quantitative extension of LTL was introduced in \cite{LMP10b}, where LTL is extended with counting quantification.  \cite{LMP10b} shows that satisfiability and model checking problems of this extension are both EXSPACE-complete. 

In \cite{GPSS80} the logic \emph{Propositional Temporal Logic (PTL)} was introduced (over discrete time models with $X$ (`next') and $U$ (`until') operators. \cite{GPSS80} shows that PTL is decidable, and it provides a sound and complete axiomatization. In \cite{SC85} it was found that the satisfiability problem for PTL is PSPACE-complete. An automata theoretic technique for obtaining satisfiability can be found in \cite{SVW85}. 

 \cite{SC85} provides a complete axiomatic system for PTL. Among the proof systems existing in literature are Hilbert-style proof system \cite{Lad87}, a Gentzen-style proof system \cite{Sza95} and a clausal resolution approach \cite{Fis91,FDP01}. These proof systems are all sound and complete.  In \cite{LPZ85} PTL was extended with the past operators, and a complete proof system for both future and past operators was presented (A detailed discussion can be found in \cite{Sza95,LP00}).   In \cite{LP00} an EXPTIME tableau algorithm is presented for the satisfiability problem of PTL.

In the literature several examples of properties of programs expressible by means of temporal logics can be found \cite{Kro87,MP81a,MP81b}. Some important properties are expressed in PTL as follows \cite{Sza95}:

\begin{itemize}
\item $p\rightarrow \Box q$: $q$ holds at all states after $p$ holds.

\item  $\Box ((\neg q) \vee (\neg p))$: $p$ and $q$ cannot hold at the same time.

\item $p\rightarrow \diamondsuit q$: $q$ holds at some time after $p$ holds. 

\item $\Box \diamondsuit p \rightarrow \diamondsuit q$: If $p$ repeatedly holds, $q$ holds after some time.

\item $\Box p \rightarrow \diamondsuit q$: If $p$ always holds, $q$ holds after some time.
\end{itemize}

Recently, more results have been presented on PTL. \cite{Rey03} showed that the satisfiability problem for PTL with the strict `until' operator is PSPACE-complete. \cite{LWW07} extended the `since-until' logic of real-line with the operators ``sometime within $n$ time units'', and they showed that the new logic is PSPACE-complete. \cite{Rey10} showed that satisfiability problem for the logic with `since-until' operators over real-numbers time is PSPACE-complete.

\section{First-Order Temporal Logics}

First-order temporal logics (FOTL) are extensions of propositional temporal logics. In addition to all propositional features these logics also allow arbitrary data structures and quantifiers. FOTLs have been extensively used in many areas including  specification and verification of reactive systems, and analysis of hardware components. First-order logics provide an  expressive formal framework for formalising the semantics of executable modal logics. They allow to obtain more robust techniques for reasoning about knowledge \cite{FHMV96,HWZ00}. FOTLs have also found applications in information systems. For example,  temporal database query languages are mainly based on first-order like languages \cite{CT98}.

Although first-order temporal logics have proved to be useful in various areas, they suffer from high computational complexity because these logics are very expressive. Indeed, most of FOTLs are not even recursively enumerable \cite{Aba89,ANS79,GHR94}. Some axiomatisations of first-order temporal logics were studied in \cite{Rey96}. In some cases, fragments of first-order temporal logics with lower computational complexity are defined through restricted extensions to propositional temporal logics \cite{Aba89,Mer92,Cho95,Pli97}.

One important logic is \emph{monodic} fragment of first-order temporal logic defined in \cite{HWZ00}, which showed that it is an expressive logic with a feasible computational behaviour. In monodic formulas, one free variable is allowed at most. In \cite{WZ02} a finite axiomatic system is presented for the monodic fragment. In \cite{ANvB98,Gra99} monodic guarded decidable fragments are introduced by restricting the quantification. 

The first-order temporal logic is represented by QTL, which includes the following syntax (which does not comprise equality and function symbols): predicate symbols, variables, constants, boolean connectives, universal quantifier and temporal operators $S$ (`since') and $U$ (`until'). Let $\mathbb{T}$ be the underlying time structures assumed for QTL constitutes strict linear orders. Then, $QTL(\mathbb{T})$ denotes the first-order temporal logic of $\mathbb{T}$, and $QTL_{fin}(\mathbb{T})$ denotes the logic of $\mathbb{T}$ with \emph{finite domains}.

\subsection{Undecidable Fragments of QTL}

In the literature, it has been known that both the monadic and two-variable fragments of first-order logic are decidable \cite{BGG97}. However, the computational complexities of their temporal counterparts are different. Let $QTL^{2}$ denote the \emph{two} - \emph{variable fragment} of QTL (where every formula contains at most two variables), and $QTL^{mo}$ denote the \emph{monadic fragment} (not monodic) of QTL (where formulas contain only unary predicates).  Assume $\mathbb{T}$ be either $\{\langle \mathbb{N},<\rangle\}$ or $\{\langle \mathbb{Z},< \rangle \} .$ Then, $QTL^{2}\cap QTL^{mo}\cap QTL(\mathbb{T})$ and $QTL^{2}\cap QTL^{mo}\cap QTL_{fin}\left(F\right)$ are not recursively enumerable \cite{HWZ00}.

\subsection{Decidable Fragments of QTL}

In the undecidable fragments given above, formulas can have the following quantification types: temporal modalities, path quantifiers and domain quantification. This causes a problem that these fragments of QTL are undecidable. It is known that the three-variable fragment of first-order logic is undecidable \cite{BGG97}.

In order to preserve decidability, corresponding fragment of QTL, which is $QTL_{1}$, contains all QTL-formulas $\varphi$, whose any subformula of the form $\varphi_{1}U\varphi_{2}$ and $\varphi_{1}S\varphi_{2}$ has at most one free variable. These formulas are \emph{monodic} (not monadic) formulas. The monodic fragments of $QTL(\langle \mathbb{N},<\rangle)$ and $QTL(\langle \mathbb{Z},<\rangle)$ are recursively enumerable \cite{HWZ01}.

Let $\mathbb{T}^{'}$ be  $\{\langle \mathbb{R},<\rangle\} $ and $\mathbb{T}$ be the following classes of time structures: ``$\{\langle \mathbb{N},<\rangle\}$, $\{\langle \mathbb{\mathbb{Z}},<\rangle\}$, $\{\langle \mathbb{\mathbb{Q}},<\rangle\}$, the class of all finite strict linear orders, and any first-order-definable class of strict linear orders'' \cite{HWZ01}. \cite{HWZ00} proves that various fragments are decidable, such as  $QTL(\mathbb{T}) \cap QTL_{1}$, $QTL(\mathbb{T}) \cap QTL_{1}^{2}$, $QTL(\mathbb{T})\cap QTL_{1}^{mo}$, $QTL_{fin}(\mathbb{T}^{'}) \cap QTL^{1}$, $QTL_{fin} (\mathbb{T}^{'}) \cap QTL_{1}^{2}$ and $QTL_{fin} (\mathbb{T}^{'})\cap QTL_{1}^{mo}$. They also provide some \emph{guarded} fragment of first-order language (For a detailed discussion, see \cite{HWZ00}).

In \cite{GKWZ02} it is shown that $QTL\left(\left\langle \mathbb{N},<\right\rangle \right)\cap QTL_{1}$ is EXPSPACE-hard. It is also shown that ``the satisfiability problem for $QTL_{1}^{mo}$-formulas in models based on $\left\langle \mathbb{N},<\right\rangle$ is EXPSPACE-complete'' \cite{HWZ01}.

Here we assumed that QTL and its fragments do not include \emph{equality} and \emph{function symbols}. It can be shown that undecidability is a major problem with the logic extended with function symbols \cite{WZ02}. For instance, ``the set of one-variable formulas with one function symbol that are valid in models based on $\left\langle \mathbb{N},<\right\rangle $ is not recursively enumerable'' \cite{HWZ01}. Moreover, ``the set of monodic QTL formulas with equality that are valid in all temporal models based on $\left\langle \mathbb{N},<\right\rangle $ is not recursively enumerable'' \cite{HWZ01}. \cite{DFL02} shows that the problem persists even for a simpler fragment. Namely, the authors prove that a fragment with ``monodic monadic two-variable formulas'' is not recursively enumerable. In \cite{WZ02} a finite \emph{Hilbert-style} axiomatisation of monodic fragment of first-order temporal logic was presented. It was also proved that ``the monodic fragment with equality is not recursively axiomatisable'' \cite{WZ02}.

Recent research results have showed that relatively expressive subsets of first-order temporal logic could be found. In \cite{WZ99,HWZ00,WZ02} suggest that expressive power of monodic first-order temporal logic can be extended further.  For example, temporal operators can be applied to formulas with more than one free variable \cite{Pli97}.  The decidability results can be also be extended to temporalties description logics. Recently, tableau-based methods  are presented for the satisfiability checking of temporal description logics  \cite{LSWZ02}. Tableau-based methods can also be devised for the satisfiability checking of decidable monodic temporal logics. This can be done by extending the tableau methods for the propositional temporal logics to the first-order case \cite{LSWZ02}. An alternative approach is to use the resolution method. \cite {DFKL08} introduces some resolution systems for monodic first-order temporal logics.

\section{Branching Time Temporal Logics}

A temporal logic system is called \emph{branching time logic} if the underlying semantics of the structure of time is branching. The underlying structure of time in branching time logics is a tree-like structure. That is, every time instant can be followed by several immediate successor time instants. In branching time logics, there are two kinds of formulas: \emph{state} formulas and \emph{path} formulas. State formulas are interpreted over states and path formulas, containing all state formulas, are interpreted over paths.

Temporal logics with underlying branching time have found many applications in artificial intelligence study. In particular, they are very useful in planning systems, where agents formulate different plans and action strategies according to different future world states \cite{McD82,RG93}. 

Since very efficient model checking algorithms have been introduced for branching time logics, these logics have been extensively used to verify finite state systems. On the other hand, in linear time logics deductive proof systems are introduced for the verification of  infinite state systems \cite{Ost92}.

An initial work about branching time logics was done by \cite{Abr80}. Later, the unified branching time system (UB) was introduced in \cite{BMP81}. A simple branching time logic, CTL, was introduced in \cite{CE82}. Thereafter, CTL* was introduced in \cite{EH86}. CTL* is an extension over CTL by adding the properties of linear time temporal logic. CTL*[P], an extension over CTL*, was introduced in \cite{LS95}. UB, CTL and CTL* include only future time temporal connectives. Whereas, CTL*[P] contains both past and future time temporal connectives.

\subsection{Computational Tree Logic (CTL)}

CTL is a point-based branching time logic, which is an extension of the logic UB by adding the operator $U$ (until).  Time is included implicitly within the temporal operators. CTL allows quantification over paths. 


Some CTL formulas are given below \cite{CGP00}:

\begin{itemize}
\item $\exists \diamondsuit(p \wedge \neg q)$: There exists a state where $p$ holds but $q$ does not hold.
\item $\forall \Box(p \rightarrow \forall \diamondsuit q)$: Whenever $p$ holds, eventually $q$ holds.
\item $\forall \Box(\exists \diamondsuit p)$: At all paths $p$ holds after some time.
\end{itemize}


CTL is a decidable logic \cite{EH82}. It can be shown that CTL has the finite model property. That is, a satisfiable formula is satisfied in a finite model of size which is bounded by ``some function of the length of the formula'' \cite{ES89}. \cite{EC82} presents a tableau method for checking the satisfiability of CTL formulas. The complexity of this procedure is EXPTIME. Model checking problem of CTL is easier than the satisfiability problem. Indeed, model checking in CTL is linear in the size of the model and the formula \cite{CES86}. This shows that model checking in CTL can be achieved very efficiently.  In \cite{Pen95} a sound and complete axiomatic system is provided for CTL. 

Recently, a quantitative extension of CTL was introduced in \cite{LMP10a}, where CTL is extended with counting quantification.  \cite{LMP10a} also provides an analysis of the expressiveness and the complexity of the model-checking problem for a range of quantitative extensions. Depending on the extension, different complexity results are obtained. 

\subsection{Full Computational Tree Logic (CTL*)}

The logic CTL* was introduced in \cite{EH86}. CTL* is an extension over CTL by adding the properties of linear time temporal logic. That is, CTL* is a logic which unifies CTL and LTL. CTL* is a more expressive logic than CTL, which makes theoretical analyses more difficult. Although model checking for CTL is linear, CTL* model checking is PSPACE-complete \cite{CES86}. Also, solving the satisfiability problem for CTL* is more difficult than solving the CTL satisfiability.  \cite{EJ00} provides an algorithm for the satisfiability problem of  CTL*, which has 2-EXPTIME complexity in the length of the formula. A sound and complete axiomatisation for CTL* has recently been defined by Reynolds in \cite{Rey01}.

\subsection{Full Computational Tree Logic with Past (CTL*[P])}

In the logics CTL and CTL* we assumed that temporal operators are restricted to future time. \cite{LS95} introduces a logic CTL*[P], which also includes past time operators. As in the linear case, addition of past operators to the language does not increase expressive power if we have a finite past; but this allows to express useful properties.  CTL*[P] is a decidable logic, which can be easily observed from the decidability of CTL*. Until recently the axiomatizability of CTL*[P] has been a long-lasting open question. \cite{Rey05} gives a sound and complete axiomatisation system for CTL*[P].

\subsection{Expressiveness of Branching Temporal Logics}

One of the main reasons of using branching time logics  is that the model-checking procedure is very efficient. The model checking task is simply to check whether a given a model satisfies a specification. CTL language allows to express useful system properties. Although model checking is expensive in linear time logics (for example, it is exponential for LTL), model checking complexity of CTL is very efficient, which is linear in the size of model and formula. However, model checking complexity of CTL* is higher than than of CTL, which is PSPACE-complete. The high complexity results from the recursive checking of all paths \cite{GRF00}.

The branching logic systems can also be used to specify properties of concurrent programs; below are some example properties expressed in UB \cite{Pen95}:

\begin{itemize}
\item $\forall \square p$: \emph{safety property}: $p$ is true at all states of each path.

\item $\forall \diamondsuit p$: \emph{liveness property}: $p$ is true at some state of each path.

\item $\exists \diamondsuit p$: \emph{possibility property}: $p$ is true at some state of some path.
\end{itemize}

Since CTL is an extension of UB, CTL subsumes the language of UB, and it can therefore express all properties which are specified in UB. CTL can express more properties such as relative ordering of events using the modality $U$ \cite{Pen95}. Both UB and CTL cannot express fairness constraints. CTL* has a more rich syntax than UB and CTL. This logic can be used in specification of more complex properties, which cannot be expressed either UB and CTL. Some examples are given below \cite{Pen95}:

\begin{itemize}

\item $\square \diamondsuit p \rightarrow \square \diamondsuit q$: \emph{fairness property}

\item $\diamondsuit \square p \rightarrow \square \diamondsuit q$: \emph{justice property}

\item $\exists\left(\left(pUq\right)\vee\square p\right)$: \emph{weak until property} 
\end{itemize}

As seen above, different combinations of linear time operators result in more expressive power. This rich syntax enables to express more complex properties, such as fairness. The branching logics mentioned in this section can be made more expressive, while still keeping all their formulas as state formulas, by allowing classical operators between the temporal and path operators. If we add past operators,  expressiveness does not increase; but the resulting logic allows more convenient notation to express some useful properties. Due to complexity and expressiveness considerations some other logics have been defined, such as CTL$^{+}$ \cite{EH82}, ECTL \cite{EC82}, ECTL$^{+}$ \cite{EC82}.

\section{Partial-Order Temporal Logics }

In concurrent systems computations are generally viewed as partially ordered sets. Since linear temporal logics are more suitable for totally ordered sets, it is difficult to apply them to concurrent and distributed systems \cite{Pr82}. Partial-order temporal logics are suitable to express partial orderings representing the behaviour of concurrent systems \cite{PW84}. Partial order structures are similar to branching structures except that each time instant can be preceded by several previous time instants. 

Initial attempts to define a logic based on partial orders were done in \cite{PW84}, where the logic POTL is introduced. POTL can express partially ordered computations without making any translation from totally orders sets. POTL can be considered as extension of the logic UB with past modalities. By adding `backward' operators POTL allows quantification over backward paths, and it can express statements requiring states with several successors and predecessors.  However, POTL framework is different than that of UB in the sense that a UB structure represents the ``set of possible computations of a program'' (where each computation is a totally ordered set); but a POTL structure represents a single computation (which is a partially ordered set)  \cite{PW84}. 
A POTL formula $q \rightarrow \forall \overleftarrow {\diamondsuit} p$ expresses that for all runs and backward fullpaths ending at states where $q$ holds, there is a state where $p$ holds \cite{Pen95}. 

\cite{PW84} shows that POTL does not have the finite model property due to the addition of past operators; but in spite of this negative result the authors show that the logic has an exponential decision procedure, and a complete axiomatisation system. 

\cite{KP86} introduces the logic POTL[U,S], which extends POTL with `until' and `since' operators. Similar to the case of POTL, POTL[U,S]   can be seen as an extension of CTL with past modalities. POTL[U,S] can express all properties POTL can express as well as the properties concerning the ``relative order of events in the future and past'' \cite{Pen95}.

Since POTL[U,S] is an extension of POTL, it does not have the finite model property. However, \cite{KP86} presents an exponential decision procedure.  A sound and complete axiomatisation system for POTL[U,S] is also given in \cite{KP86}. POTL[U,S] has a high model checking complexity, because formulas contain past modalities, and they are ``interpreted over models corresponding to runs of concurrent systems'' \cite{Pen95}. Indeed, \cite{KP86} shows that the complexity is exponential in the model size and doubly exponential in the formulas size. 

In literature, there are more recent results for  logics with partial-order semantics. \cite{BP97} presents a new temporal logic, where linear and partial order semantics are combined. Namely, a computation is modeled as a linear sequence of states, which are associated with ``past partial-order history''. The authors also give a sound and partially complete proof system for the logic. In \cite{GKP02} partial order reductions are studied for the logics CTL and CTL*  based on the partial order techniques to reduce the state space. \cite{AR04} introduces a new partial-order temporal logic based on different semantical model to increase the expressiveness. In \cite{LP10} partial-order reduction techniques are applied to  linear and branching time temporal logics for knowledge (without the next operator) to reduce the model size before applying model checking procedure.

\section{Interval Temporal Logics} \label{sec:Interval-Temporal-Logics}

\emph{Interval temporal logics} are temporal logics which allow reason about periods of time. Since representation of logical reasoning about periods are more expressive than reasoning about points, the interval-based scheme provides us with a richer representation formalism than the point-based approach.

In this section, we present a selection of well-known interval temporal logics. In literature, many similar logics can be found; but most of these logics are generalisations or specialisations of the ones we will discuss below. 

\subsection{Propositional Interval Temporal Logics }

In this section we will present the well-known propositional interval logics, which involve unary or binary modal operators, and whose semantic structures are over partial orderings with linear interval property, i.e. ``orderings in which every interval is linear'' \cite{GMS04}.

The syntax of propositional interval temporal logics is constructed from the following: the set of propositional variables, the truth values, the classical operators (boolean operators, negation, etc.), and a set of temporal operators defined for each logic.

\subsubsection{The Logic HS }

The logic HS \cite{HS91} is a relatively expressive propositional interval temporal logic. All modal operators of HS are unary. The logic HS has enough expressive power to distinguish different temporal structures, such as of discrete, continuous, bound, linear or complete time structures.  These are formally shown as follows \cite{HS91}:

\begin{itemize}
\item $length0$ $\equiv [B]\bot$
\item $length1$ $\equiv \langle B \rangle\top \wedge [B] length0$ ($length1$ holds at intervals with no proper subintervals.)
\item $dense$  $\equiv \neg length1$
\item $discrete$ $\equiv$ $length0$  $\vee$  $length1$ $\vee$  ($\langle B \rangle length1$  $\wedge$  $\langle E \rangle length1$)
\end{itemize}

\noindent where $\langle B \rangle \phi$ is true  iff $\phi$ holds at some interval that begins with the current interval and ends before it ends; $\langle E \rangle \phi$ is true iff  $\phi$ holds at some interval that begins after the current interval starts and ends when it ends; and  $[B]\phi$ is defined as $\neg\langle B \rangle \neg\phi$.

HS is a quite expressive logic due to its large modal operator set. However, it is not axiomatisable and is highly undecidable \cite{HS91}. The following theorems are taken from \cite{HS91}:

\begin{itemize}
\item ``The validity problem interpreted over any class of ordered structures
with an infinitely ascending sequence is r.e.-hard (Thus, in particular,
HS is undecidable for the class of all (non-strict) models, linear
models, discrete linear models, dense linear models and unbounded
linear models).''
\item ``The validity problem interpreted over any class of Dedekind complete
ordered structures having an infinitely ascending sequence is ${\scriptstyle \prod_{1}^{1}}$-hard
(For instance, the validity in any of the orderings of the natural
numbers, integers, or reals is not recursively axiomatisable. Undecidability
even occurs in the classes of structures with no infinitely ascending
sequences).''
\item ``The validity problem interpreted over any class of Dedekind complete
ordered structures having unboundedly ascending sequences is co-r.e.-hard.''
\end{itemize}

Undecidability results given above are based on the observation that HS formulas encode the computation of a Turing machine. In \cite{MR99} undecidability was proved by means of tiling problem.

In \cite{MV97} some interesting results for the logic HS were presented. By using a geometrical representation for the modalities a sound and complete proof system for HS was introduced. \cite{MV97}  also proved that HS is a more expressive logic than any other temporal logic based on ``linear orderings of time instants''.

In \cite{HS91} a translation machinery that converts an HS formula to its equivalent first-order formula on a corresponding first-order structure was provided. Such a translation is useful to reduce problems to well-known results in first-order logic.

\subsubsection{The Logic CDT }

The logic CDT was introduced by Venema in \cite{Ven91}. It is one of the most expressive propositional interval logic over linear orderings \cite{GMS04}. CDT includes the binary modal operators $C,D$ and $T$. These operators subsume all unary modalities of propositional interval logics of Allen's interval relations \cite{Bre07}. CDT can distinguish different classes of temporal structures, such as of discrete, continuous, bound, linear or complete time structures. For example, an interval's being discrete can be specified in CDT as follows:

\begin{itemize}
\item ($length1$ $C$ $\top) \wedge (\top$ $C$ $length1$).
\end{itemize}

\cite{Ven91} gives an axiomatic system which is sound and complete for the logic CDT which is interpreted over non-strict linear models. This axiomatic system can be extended for the classes of discrete linear orderings, dense linear orderings, etc. \cite{GMS04}. Since CDT subsumes HS, the satisfiability problem for ``CDT is not decidable over almost all classes of linear orderings, including discrete, dense, continuous, etc.'' \cite{GMS04}.

The partial order semantics of CDT has been recently studied in \cite{GMS03b}, where the logic BCDT$^{+}$ is introduced. BCDT$^{+}$ uses the language of CDT with partial order semantics of linear intervals. To our best knowledge the decidability and axiomatizability of the strict versions of CDT and BCDT$^{+}$ are still open.

\subsubsection{The Logic PNL}

Propositional Neighbourhood Logic (PNL) is the propositional fragment of First-Order Neighbourhood Logic introduced in \cite{CH98}. It has been studied on both strict and non-strict linear structures in \cite{GMS03a}. The language with non-strict semantics is called PNL$^{\pi+}$ including the modalities $\diamondsuit_{r}$ (\emph{met by}) and $\diamondsuit_{l}$ (\emph{meets}), and the model constant $\pi$. The modal operators can have either strict or non-strict semantics.

Assume PNL$^{+}$ denotes the non-strict PNL without the modal constant $\pi$, and PNL$^{-}$ denotes the strict PNL without the modal constant $\pi$ . The logic PNL$^{\pi+}$ subsumes both PNL$^{+}$ and PNL$^{-}$  \cite{Bre07}.

Given that formulas are interpreted over strict linear models, PNL$^{-}$ has enough expressive power to distinguish the different classes of linear structures, such as discreteness, continuity, boundness, or completeness. For example, unboundness and density can be specified in PNL$^{-}$ as follows \cite{GMS04}:

\begin{itemize}
\item \emph{unbound} $\equiv \Box_{r} \phi \rightarrow \diamondsuit_{r} \phi$
\item \emph{dense} $\equiv (\diamondsuit_{r} \diamondsuit_{r} \phi \rightarrow \diamondsuit_{r} \diamondsuit_{r} \diamondsuit_{r} \phi) \wedge (\diamondsuit_{r} \Box_{r} \phi \rightarrow \diamondsuit_{r} \diamondsuit_{r} \Box_{r} \phi)$
\end{itemize}

In \cite{GMS03a} several sound and complete axiomatic systems were provided for various classes of models. In addition to strict linear models \cite{GMS03a} also provides sound and complete axiomatic systems for non-strict linear structures, complete unbounded linear structures, unbounded structures, dense structures, discrete structures, dense unbounded structures and discrete unbounded structures. As for decidability results, \cite{BMS07a} shows that the satisfiability problem for PNL$^{\pi+}$, PNL$^{+}$ and PNL$^{-}$ over the integers is NEXPTIME-complete. \cite{BMS07a}  introduces a sound and complete tableau algorithm, and shows that it is optimal. In \cite{BGMS07}, the expressive power of PNL$^{\pi+}$, PNL$^{+}$ and PNL$^{-}$ is compared, and it is shown that PNL$^{\pi+}$ is strictly more expressive than PNL$^{+}$ and PNL$^{-}$. \cite{BGMS07} proves that ``the satisfiability problem for PNL$^{\pi+}$ over the class of all linear orders, as well as over some natural subclasses of it, such as the class of all well-orders and the class of all finite linear orders, can be decided in NEXPTIME by reducing it to the satisfiability problem for the two-variable fragment of first-order logic over the same classes of structures''.

An important fragment of the PNL is the \emph{Right Propositional Neighbourhood Logic} (RPNL) which is based on the right neighbourhood relation between intervals. The language with non-strict semantics is called RPNL$^{\pi+}$. The non-strict fragment without the modal constant $\pi$ is denoted by RPNL$^{+}$, an the strict fragment without the modal constant $\pi$ is denoted by RPNL$^{-}$. As for decidability results, in \cite{BM05} an EXSPACE tableau-based decision procedure is devised for RPNL$^{-}$ interpreted over natural numbers. In \cite{BMS07b} another  NEXPTIME decision procedure  is developed. This method works for all classes of RPNL, which are RPNL$^{\pi+}$, RPNL$^{+}$, and RPNL$^{-}$, interpreted over natural numbers. \cite{BMS07b} also proves the optimality of the decision procedure.

\subsubsection{Subinterval Logics} For many years, the high computational complexity of interval logics (such as HS and CDT) restricted these logics in practical applications and semantic investigation. Recently, the trend has shifted to finding expressive decidable fragments. The most important decidable fragments are PNL and its fragments, logics of neighbourhood \cite{BGM09}, \emph{sub-interval} and \emph{superinterval} structures \cite{BGM10}. In \cite{BGM10} the logics of subinterval structures over \emph{dense} linear orders is shown to be decidable. \cite{BGM10} also provides a tableau-based decision procedure, which is shown to be PSPACE-complete.  \cite{MPS10} shows that ``the satisfiability problem for interval logics of the reflexive sub-interval and super-interval relations interpreted over \emph{finite} linear orders is PSPACE-complete''. 

Subinterval logics have also been investigated in natural language discourse. In \cite{Pra05} a sub-interval logic, which is used in capturing temporal prepositions of a natural language, is introduced. In \cite{Kon06} a quantitative extension of this logic is represented. Both logics are decidable, and their satisfiability problems are in NEXPTIME.

\subsection{First-Order Interval Temporal Logics }

First-order interval temporal logics were originally defined to formally specify and verify hardware components of real-time systems.  ITL is the most commonly known first-order interval temporal logic. Numerous extensions of ITL, such as Duration Calculus \cite{CHR91}, Neighbourhood Logic \cite{CHR91} etc., have been introduced. Below we will review well-known first-order interval temporal logics. 

\subsubsection{\label{sub:The-Logic-ITL}The Logic ITL }

ITL was first introduced in \cite{Mos83} (which was ``interpreted over discrete linear orderings with finite time intervals'' \cite{GMS04}). The
formulas of ITL are constructed from the following: an infinite set of global (independent of time and time intervals) variables, an infinite set of temporal variables, an infinite set of global function symbols, an infinite set of predicate symbols and an infinite set of temporal propositional letters.

Not surprisingly ITL is highly undecidable. A sound and complete axiomatic system is represented in \cite{Dut95}.  
\cite{Dut95,Mos00b} consider some local variants of ITL, and provide sound and complete proof systems for ITL with locality constraint. \cite{Gue00} provides a complete proof system for  ITL extended with projection. \cite{Gue98} studies probabilistic interval temporal logic. 

\subsubsection{The Logic NL}

Although ITL is a very expressive logic, it has a limitation that it does not allow to reason about outside of the current interval. The logic NL, proposed in \cite{CHR91}, solves this problem. Its left \emph{neighbourhood modality} $\diamondsuit_{l}$ and right \emph{neighbourhood} \emph{modality} $\diamondsuit_{r}$ can allow  to look outside of the interval.

NL can express any of the Allen's interval relations; thus, it can represent important properties, such as discreteness, density, boundedness, etc; for example, the chop operator $C$ can be expressed in terms of the modalities $\diamondsuit_{l}$ and $\diamondsuit_{r}$ as follows \cite{GMS04}:

\begin{itemize}
\item $\phi C\psi=\exists x,y\left(\ell=x+y\right)\wedge\diamondsuit_{l}\diamondsuit_{r}\left(\left(\ell=x\right)\wedge\phi\wedge\diamondsuit_{r}\left(\left(\ell=y\right)\wedge\psi\right)\right)$
\end{itemize}

NL is an undecidable logic like ITL. In \cite{BRC00} a sound and complete axiomatic system is given for the logic NL. In \cite{BC97} \emph{up} and \emph{down} modalities, represented by $\diamondsuit_{u}$, $\diamondsuit_{d}$ respectively, were introduced, and two dimensional version of NL was proposed.

\subsubsection{The Logic DC }

Duration Calculus (DC) \cite{CHR91} is a first-order interval temporal logic with the additional notion of \emph{state}, which is characterised by a \emph{duration}\footnote{``The duration of a state is the length of the time period during which the system remains in the state'' \cite{GMS04}.}. DC is an extension of ITL that temporal variables other than $\ell$ have a  \emph{state expression} structure. The special interval variable $\ell$ denotes the interval \emph{length}.

DC has been used in the specification and verification of various complex systems. As a specification example, we specify the real-time requirement of a gas burner system, which is ``the proportion of leak time in an interval is not more than one-twentieth of the interval, if the interval is at least one minute long'', which is expressed in DC as follows \cite{CHR91}:

\begin{itemize}
\item Req $\equiv \ell\geq60$ $\Rightarrow$ 20$\int Leak\leq\ell$
\end{itemize}

All axioms and inference rules of ITL can be adopted in DC. However, additional axioms are needed for temporal variables. In \cite{CH04} an axiomatic system for Duration Calculus is given. The satisfiability problem for both first-order and propositional DC is shown to be undecidable \cite{CHS93}.

Several fragments of DC have been investigated so far. In \cite{CHS93} a fragment of propositional DC, called RDC, was introduced. It was shown that RDC has a decidable satisfiability problem when interpreted over $\mathbb{N}$, $\mathbb{Q}$ and $\mathbb{R}$. In \cite{Rab98} the satisfiability problems of several extensions of RDC were studied . In \cite{Fra96} an extension of RDC was presented on continuous time ``with a restriction on the finite variability such that the number of discontinuous points of any state in any unit interval has a fixed upper bound''. In \cite{GD99} a decidable variant of DC was presented, where negation is removed from the syntax; but an iteration operator is introduced together with some form of inequalities. In \cite{CSFDC00}  another fragment of propositional DC, which can capture Allen's relations \cite{All83},  was introduced by imposing some syntactic restrictions.  By proposing a sound, complete and terminating decision algorithm, it was shown that the satisfiability problem is decidable. In \cite{Pan01} a logic with quantification over states was introduced. It was shown that  the satisfiability of formulas is decidable. This decision algorithm was implemented as a tool called DCVALID. In \cite{CH98} Duration Calculus and first-order neighbourhood logic were combined, and a axiomatic systems for DC and NL were merged. It was proved that ``the fragment of DC/NL obtained by restricting the formulas'' of state expressions  is decidable \cite{GMS04}.  An extension with formulas with equality becomes undecidable. 

Model checking problem for DC is a challenging task. In general, there has not been a general model checking technique for this logic. To have efficient model checking techniques, it is necessary to consider a fragment of the logic. In \cite{Fra96} some model checking tools were developed for a class of models which are restricted to some possible behaviours of real-time systems. In \cite{Zho94,DH96,TH98,TH04} some techniques were developed to check if a timed automaton satisfies a formula of the type ``linear duration invariants''. In \cite{SHP98} some algorithms were developed to check the satisfiability over integer models.  In \cite{Pan01} a DC validity checker, called DCVALID, to check the satisfiability of formulas which are interpreted over discrete-time. \cite{Fra02} suggested \emph{bounded validity checking} \cite{BCCZ99} of ``a discrete-time DC without timing constraints by polynomial-sized reduction to propositional SAT solving''. In \cite{FH07} a decidability result  and a model-checking algorithm are presented ``for a rich subset of DC through reductions to first-order logic over the real-closed field and to multi-priced timed automata (MPTA)''. 

\subsubsection{The Logic IDL}

Duration Calculus is  a very expressive logic for specifying real-time requirements; but the automata theory for DC models is rather primitive and there are no available tools. By contrast, the state sequences with time has been widely used in  real-time system behaviour \cite{AD96}. The automata theory of timed state sequences have been applied to tools such as Hytec \cite{AHH96}, Uppaal \cite{BLLPY96}, Kronos \cite{BDMOTY98} etc. 

\cite{Pan02} introduced \emph{Interval Duration Logic (IDL)}, which is defined on \emph{timed state sequence} models and incorporates formulas with \emph{cumulative} amount of time. Due to its expressive syntax it can express complex real-time properties, e.g. scheduling and planning constraints.  As an example, we give a specification example from a gas burner system. The property `between two instances of Leak there is at least \emph{k} seconds' is specified in IDL as follows:

\begin{itemize}
\item $\Box((\lceil\lceil Leak \rceil^{\frown}\lceil\lceil \neg Leak \rceil^{\frown}\lceil\lceil Leak \rceil^{0}) \Rightarrow \ell \geq k)$
\end{itemize}

IDL is a very expressive logic; but it is undecidable. However, there are some methods which have been proposed for the satisfiability and model checking problems of IDL. \cite{SPS05} applies \emph{bounded validity checking} technique \cite{BCCZ99} to IDL  ``by polynomially reducing this to checking unsatisfiability of \emph{lin-sat} formulae''. \cite{SPS05} also compares various methods for the satisfiability problem ``including digitization technique \cite{CP03}, combined with an automata-theoretic analysis \cite{Pan01}; digitization technique \cite{CP03} followed by pure propositional SAT solving \cite{Fra02}; and (c) lin-sat solving \cite{FORS02}''.

\cite{Pan02} presents a decidable subset of IDL, which has a restriction that only \emph{located time constraints} are allowed. The paper shows that the models of this subset can be considered as ``timed words accepted by a finite state event-recording integrator automaton'', which implies the satisfiability of the subset. It is also shown that the defined subset and event-recording automata have the same expressive power, which makes this logic an important decidable subset in the domain of DC.

\section{Conclusion} \label{sec:conclusion}

In this survey paper we have outlined recent important developments
on propositional/first-order linear temporal logics, branching temporal logics, partial-order temporal logics and interval temporal logics by presenting important features, such as (un)decidability results, expressiveness and axiomatization systems. For a comparison of features of the temporal logics we discussed see Table 1. Note that we use the following abbreviations: \emph{No*}: Undecidable in general, but decidable for some fragments or specific cases; \emph{No**}: No deduction system in general, but available for some fragments or specific cases; \emph{No***}: No model checking algorithm in general, but available for some fragments or specific cases; \emph{Yes*}: Decidable for some time domains; \emph{Yes**}:  Available for some time domains; \emph{Yes***}: Available for some time domains.

\begin{landscape}
\begin{table}
\caption{A comparison of features of temporal logics.}
\centering
\small
\begin{tabular}{|c|c|c|c|c|c|c|c|}
\hline
\textbf{Logic} & \textbf{Logic Order} & \textbf{Fund. Entity} & \textbf{Temp. Struc.} &\textbf{Metric for Time} & \textbf{Decidability} & \textbf{Deductive Sys.} & \textbf{Model Checking}\tabularnewline\hline\hline
LTL&Propositional&Point&Linear&No&Yes&Yes&Yes\tabularnewline\hline
PTL&Propositional&Point&Linear&No&Yes&Yes&Yes\tabularnewline\hline\hline
QTL&First-order&Point&Linear&No&No*&No**&?\tabularnewline\hline\hline
CTL&Propositional&Point&Branching&No&Yes&Yes&Yes\tabularnewline\hline
CTL*&Propositional&Point&Branching&No&Yes&Yes&Yes\tabularnewline\hline
CTL*[P]&Propositional&Point&Branching&No&Yes&Yes&Yes\tabularnewline\hline
POTL&Propositional&Point&Partial&No&Yes&Yes&?\tabularnewline\hline\hline
HS&Propositional&Interval&Linear&No&No&No&No\tabularnewline\hline
CDT&Propositional&Interval&Linear&No&No&Yes&No\tabularnewline\hline
PNL&Propositional&Interval&Linear&No&Yes&Yes&No\tabularnewline\hline
ITL&First-order&Interval&Linear&No&No&Yes&No\tabularnewline\hline
NL&First-order&Interval&Linear&Yes&No*&Yes&No\tabularnewline\hline
DC&First-order&Interval&Linear&Yes&No*&Yes&No***\tabularnewline\hline
IDL&First-order&Interval&Linear&Yes&No*&No&No***\tabularnewline\hline\hline
\end{tabular}
\end{table}
\end{landscape}

\subsection*{Acknowledgements.}
This work was partially supported by EPSRC research project EP/F033567.

\bibliographystyle{plain}       
\bibliography{bibliography}     

\end{document}